\documentclass[sigconf,screen]{acmart}

\usepackage[linesnumbered,ruled,vlined]{algorithm2e}
\usepackage{subfigure}
\usepackage{multirow}
\usepackage{multicol}
\usepackage{tcolorbox}
\usepackage{listings}
\usepackage[T1]{fontenc}
\usepackage{aecompl}

\AtBeginDocument{
  }

\setcopyright{rightsretained}
\acmPrice{}
\acmDOI{10.1145/3611643.3613869}
\acmYear{2023}
\copyrightyear{2023}
\acmSubmissionID{fse23industry-p25-p}
\acmISBN{979-8-4007-0327-0/23/12}
\acmConference[ESEC/FSE '23]{Proceedings of the 31st ACM Joint European Software Engineering Conference and Symposium on the Foundations of Software Engineering}{December 3--9, 2023}{San Francisco, CA, USA}
\acmBooktitle{Proceedings of the 31st ACM Joint European Software Engineering Conference and Symposium on the Foundations of Software Engineering (ESEC/FSE '23), December 3--9, 2023, San Francisco, CA, USA}

\begin{document}

\title{xASTNN: Improved Code Representations for Industrial Practice}

\author{Zhiwei Xu}
\affiliation{
  \institution{KLISS, BNRist, School of Software}
  \institution{Tsinghua University}
  \city{Beijing}
  \country{China}
}

\author{Min Zhou}
\authornote{Min Zhou is the corresponding author (mzhou@tsinghua.edu.cn).}
\affiliation{
  \institution{KLISS, BNRist, School of Software}
  \institution{Tsinghua University}
  \city{Beijing}
  \country{China}
}

\author{Xibin Zhao}
\affiliation{
  \institution{KLISS, BNRist, School of Software}
  \institution{Tsinghua University}
  \city{Beijing}
  \country{China}
}

\author{Yang Chen}
\affiliation{
  \institution{School of Computer Science}
  \institution{Fudan University}
  \city{Shanghai}
  \country{China}
}

\author{Xi Cheng}
\affiliation{
  \institution{VMware}
  \city{Beijing}
  \country{China}
}

\author{Hongyu Zhang}
\affiliation{
  \institution{School of Big Data and Software Engineering}
  \institution{Chongqing University}
  \city{Chongqing}
  \country{China}
}

\begin{abstract}
The application of deep learning techniques in software engineering becomes increasingly popular. One key problem is developing high-quality and easy-to-use source code representations for code-related tasks. The research community has acquired impressive results in recent years. However, due to the deployment difficulties and performance bottlenecks, seldom these approaches are applied to the industry. In this paper, we present xASTNN, an eXtreme Abstract Syntax Tree (AST)-based Neural Network for source code representation, aiming to push this technique to industrial practice. The proposed xASTNN has three advantages. First, xASTNN is completely based on widely-used ASTs and does not require complicated data pre-processing, making it applicable to various programming languages and practical scenarios. Second, three closely-related designs are proposed to guarantee the effectiveness of xASTNN, including statement subtree sequence for code naturalness, gated recursive unit for syntactical information, and gated recurrent unit for sequential information. Third, a dynamic batching algorithm is introduced to significantly reduce the time complexity of xASTNN. Two code comprehension downstream tasks, code classification and code clone detection, are adopted for evaluation. The results demonstrate that our xASTNN can improve the state-of-the-art while being faster than the baselines.
\end{abstract}



\maketitle

\section{Introduction}

Code representations (a.k.a., embeddings) is of great importance in deep learning-based software engineering techniques. A high-quality representation model can significantly improve the performance of many downstream tasks, such as code search \cite{gu2018deep, salza2022effectiveness, cambronero2019deep}, code clone detection \cite{wei2017supervised, fang2020functional, zhao2018deepsim}, and bug localization \cite{huo2019deep}. Recently, code representation learning has aroused much interest in both academia \cite{han2021comparison} and industry \cite{cito2022counterfactual, zlotchevski2022exploring, siow2022learning, zhou2022improving}.

Despite advancement, there are still restrictions that prevent the widespread adoption of existing code representation approaches in industry. Effectiveness, efficiency, and applicability are of particular concern. A recent study \cite{wu2020scdetector} has shown that the state-of-the-art approach ASTNN \cite{zhang2019novel} incurs much computation time to embed the source code as vectors for code clone detection (e.g., $5.48 \times$ over SCDetector \cite{wu2020scdetector} and $3.09 \times$ over RtvNN \cite{white2016deep}). Kang et al. \cite{kang2019assessing} conduct an empirical study to demonstrate that the popular code representation approach code2vec \cite{alon2019code2vec} lacks generalizability and cannot be readily leveraged for downstream tasks. GNN-CDFG \cite{brauckmann2020compiler}, GGNN \cite{allamanis2017learning}, and HPG+HGT \cite{zhang2022learning} require their elaborate graphs to represent code segments, leading to a strong dependency on the characteristics of programming languages. flow2vec \cite{sui2020flow2vec}, inst2vec \cite{ben2018neural}, and DeepSim \cite{zhao2018deepsim} are designed based on specific compilers such as LLVM \cite{lattner2004llvm} and WALA \cite{wala2006watson}. The above restrictions hinder the industrial applications of existing code representation approaches.

In the industrial practice of neural code representation, we are often involved in a variety of production issues, such as constrained computing resources, limited response time, abnormal data inputs, etc. Therefore, the code representation approaches are supposed to address the following three imperative challenges: (1) Design a model with notable effectiveness. The quality of code representations can directly influence the effectiveness of the model on downstream tasks. (2) Build a model that is as fast and lightweight as possible. It is unacceptable in the industry that the software services have severe runtime delays or memory overflow problems. (3) The model should be applicable in various scenarios, e.g., be independent of parsers to cater for different programming languages, and be able to handle code segments of various sizes so as not to be trapped in gradient problems.

To this end, we propose an eXtreme abstract syntax tree (AST)-based neural network (xASTNN) for neural code representations in industrial practice. xASTNN is entirely based on common ASTs, which are always accessible and available with the source codes. To guarantee the effectiveness of xASTNN, we first perform a preorder traversal upon the AST to convert it into a statement subtree sequence. Further, we present gated recursive unit (GRvU) for capturing the syntactical information of each subtree and gated recurrent unit (GRtU) for capturing the sequential information of the subtree sequence. The introduction of gating mechanism not only improves the generalizability of our approach, but also prevent our approach from gradient problems encountered by the previous approach \cite{zhang2019novel}. To optimize the computation time of our xASTNN, we describe a novel and high-efficient dynamic batching algorithm for the processing of subtrees, which enables parallel operations at the contents of the same tree depth within a batch of data samples. Under the above designs, our xASTNN is thus effective, efficient, and can be widely used in industrial practice.

Extensive experiments have been carried out to validate the performance of our xASTNN. Two comprehension downstream tasks including code classification and code clone detection are applied for evaluation. The results demonstrate that our xASTNN is highly competitive as a code representation model for practical usage. Specifically, our xASTNN improves the state-of-the-art performance and achieves superb efficiency at the same time. For example, our xASTNN achieves an accuracy of 0.985 for the POJ dataset in the code classification task. The computation time of our xASTNN for representing one code segment is over 10$\times$ faster than the previous approach ASTNN. 

To summarize, we have made the following contributions:

\begin{itemize}
    \item (\textit{Applicability}) Based on easily accessible ASTs, we present an innovative neural code representation approach named xASTNN for industrial practice. 
    \item (\textit{Effectiveness}, \textit{Applicability}) We introduce gating mechanism to effectively capture the syntactical and sequential information of statement subtree sequence while avoiding gradient problems encountered by the previous approach. 
    \item (\textit{Efficiency}) We describe a high-efficient dynamic batching algorithm that greatly optimizes the time complexity of operations over trees.
    \item Our representation approach is evaluated on two common comprehension tasks: code classification and code clone detection. The results demonstrate that the proposed approach can outperform the state-of-the-art and is more practical.
\end{itemize}

The remainder of the paper is organized as follows. Section \ref{sec:related_work} introduces the related work. Section \ref{sec:approach} presents the proposed approach, including statement subtree sequence, gated recursive unit, gated recurrent unit, and dynamic batching algorithm. Section \ref{sec:evaluation} validates the competing approaches. Section \ref{sec:threats} illustrates the threats to validity. Section \ref{sec:lessons_learned} discusses the lessons learned from the practice and we conclude in Section \ref{sec:conclusion}.

\section{Related Work}
\label{sec:related_work}

The emergence of code representation learning mainly borrows the concepts from natural language processing (NLP), and the source code are treated as a sequence of tokens. Raychev \textit{et al.} \cite{raychev2014code} describes a recurrent neural network (RNN) and N-gram model for code completion. Allamanis \textit{et al.} \cite{allamanis2013mining} train an N-gram model on the GitHub Java corpus for mining source code repositories. Further, they \cite{allamanis2015suggesting} propose a neural context model to suggest class and method names. CODE-NN \cite{iyer2016summarizing} exploits LSTM \cite{hochreiter1997long} and neural attention to summarize code. However, the above approaches ignore the inherent syntactical features of programming languages, thus failing to produce good code representations.

To capture syntactical information of programming languages, recent researches introduce ASTs as input to the representation models. AutoenCODE \cite{white2016deep} adopts a recursive autoencoder over ASTs to learn the unsupervised program embeddings. TBCNN \cite{mou2016convolutional} proposes a tree-based convolution over the AST to represent the code segment. CDLH \cite{wei2017supervised} introduces TreeLSTM \cite{tai2015improved} to generate the embeddings for code clone detection. Different from previous work that directly captures the syntactical information from ASTs, code2vec \cite{alon2019code2vec} first extracts a bag of leaf-to-leaf paths from the AST and aggregates these paths using the attention mechanism. code2seq \cite{alon2018code2seq} improves code2vec by introducing LSTM \cite{hochreiter1997long} to encode paths, which is applied in code summarization. A similar approach to ours is ASTNN \cite{zhang2019novel}, which encodes the source code by capturing both the lexical and syntactical knowledge of statements. Since code naturalness is well modeled, ASTNN achieves superb performance in code tasks. Nevertheless, ASTNN is proven to be inefficient when applying to practical scenarios \cite{wu2020scdetector}.

There are also many code representation approaches based on pre-training techniques. CodeBERT \cite{feng2020codebert} considers the source code as token sequence and applies the successful BERT \cite{devlin2018bert} to pre-train the model. GraphCodeBERT \cite{guo2020graphcodebert} improves CodeBERT by introducing the data flow of the source code. InferCode \cite{bui2021infercode} describes a self-supervised pre-training approach by predicting subtrees of ASTs. Mastropaolo \textit{et al.} \cite{mastropaolo2021studying} pre-train a T5 model based on a dataset composed of natural language English and source code for code-related tasks. SPT-Code \cite{niu2022spt} presents a sequence-to-sequence pre-training pipeline for source code representations. However, recent evidence \cite{karmakar2021pre} points out that existing pre-training code representation approaches may fail in some code tasks and even achieve worse performance than BERT \cite{devlin2018bert}. Besides, the pre-training techniques require a large-scale dataset and a large amount of computing resources, which can be a huge challenge for industrial practice.

Therefore, an effective, efficient, and general-purpose code embedding approach is urgently needed. We propose xASTNN, aiming to resolve the practical challenges faced by existing approaches.

\begin{figure}[t]
  \centering
  \includegraphics[width=\linewidth]{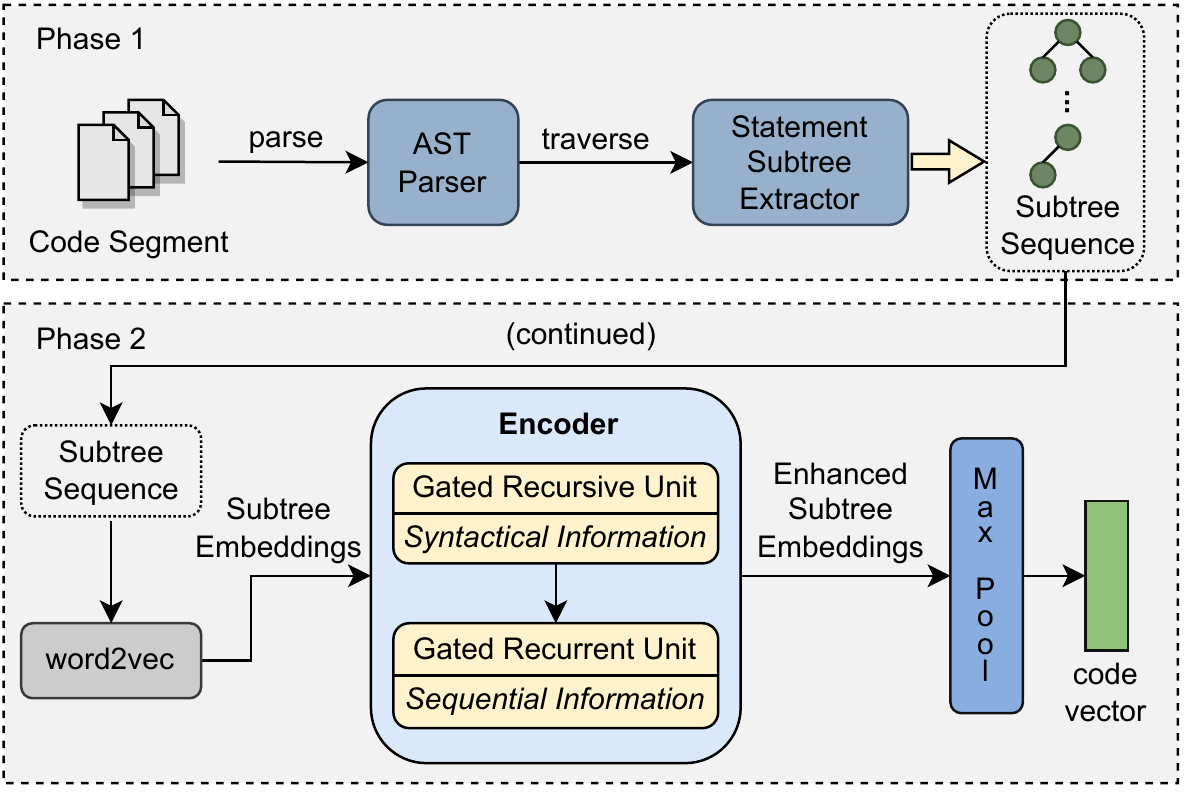}
  \caption{Overview of the proposed approach. The first phase parses the code segment into a statement subtree sequence to introduce the code naturalness. The second phase combines the syntactical information and sequential information to enhance the embeddings of the subtree sequence and leverages a max pooling layer to produce code representations.}
  \label{fig:overview}
\end{figure}

\section{Approach}
\label{sec:approach}

In this section, we present the proposed approach xASTNN. As shown in Figure \ref{fig:overview}, xASTNN consists of two phases. In the first phase, the code fragment is first transformed into an AST by using a common AST parser. A preorder traversal over AST is applied to obtain a statement subtree sequence, in which each subtree corresponds to a statement of the code fragment. The first phase can be computed in advance when training xASTNN. In the second phase, we first adopt the popular word2vec \cite{mikolov2013distributed} to embed the subtree sequence into a distributed space. To ensure the effectiveness and applicability of xASTNN, gating mechanism is introduced to incorporate the syntactical information of subtrees and sequential information of subtree sequence. Lastly, we exploit a max pooling layer to combine the subtree embeddings into the code vector. 

\subsection{Statement Subtrees for Code Naturalness}
The code naturalness is a hypothesis stating that software corpora have similar statistical properties to natural language corpora since the software is a form of human communication \cite{allamanis2018survey}. To capture the code naturalness, we first consider the source code as an AST and then transform it into a combination of statement subtrees as input to the neural network. AST is a widely used structure in the field of code representation approaches and can be easily accessed by common AST parsers. In this work, we exploit javalang \cite{javalang} and pycparser \cite{pycparser} for the experiments of Java and C, respectively. Based on our empirical study, we opt to extract the subtrees at the statement level for the following two reasons.

\begin{itemize}
    \item Previous work \cite{zhang2019novel} has shown that statement subtree sequence can effectively capture code naturalness, thereby improving the quality of code representations. Besides, subtree at the statement level is a good trade-off between the size of subtree and the richness of syntactical information.
    \item The size of statement subtrees is approximately equal in comparison to subtrees of other granularities. For example, subtree at the block level is made up of a varying number of statement subtrees, which can easily lead to an unstable size of subtree and thus introduces efficiency concerns in practice (see Section \ref{sec:batching}).
\end{itemize}

Algorithm \ref{alg:split} illustrates how we transform an AST into a statement subtree sequence. This algorithm takes the root of an AST $head$ and the set of subtree root identifiers $R$ as input, and outputs the subtree sequence extracted from the AST. The subtree root identifiers target to help the preorder traversal process to recognize statements, which generally are logical tokens in programming languages such as $if$, $while$, $for$, $switch$, $FunctionDeclaration$, $VariableDeclaration$, $FuncCall$, etc. They can be easily obtained when it comes to other programming languages.

\begin{algorithm}[t]
    \SetAlgoLined 
    \caption{Transform an AST into subtree sequence}
    \label{alg:split}
    \KwIn{Root of an AST $head$\; 
	\hspace{0.95cm} Set of subtree root identifiers $R$; \\
	}
    \KwOut{Subtree sequence $S$;}
    \BlankLine
    \SetKwFunction{FA}{preorderTraversal}
    \SetKwProg{An}{Function}{}{}
    \An{\FA{$root$}}{
        \If{$root$ in $R$}{
            s $\leftarrow$ construct a statement subtree based on $root$
            
            append $s$ to $S$
        }
        \For{\rm{each} $child$ \rm{of} $root$, $child \notin s$}{
                \texttt{preorderTraversal($child$)}
        }
    }
    \BlankLine
    Initialize $S \leftarrow []$\;
    \texttt{preorderTraversal}($head$)\;
    \KwRet $S$\;
\end{algorithm}

\begin{figure}[t]
    \hspace{0.43cm}
    \begin{minipage}[c]{0.3\linewidth}
        \begin{lstlisting}
while (i < 5) {
    if (a < 100) {
        temp = a;
        a = b;
        b = temp;
    }
    else 
        a = a % 100;
    i = i + 1;
}
        \end{lstlisting}
        {\footnotesize \vspace{0.44cm} (a) Example code segment.}
        \vspace{-0.55cm}
    \end{minipage}
    \hspace{1.5cm}
    \begin{minipage}[c]{0.4\linewidth}
        \centering
        \includegraphics[width=\linewidth]{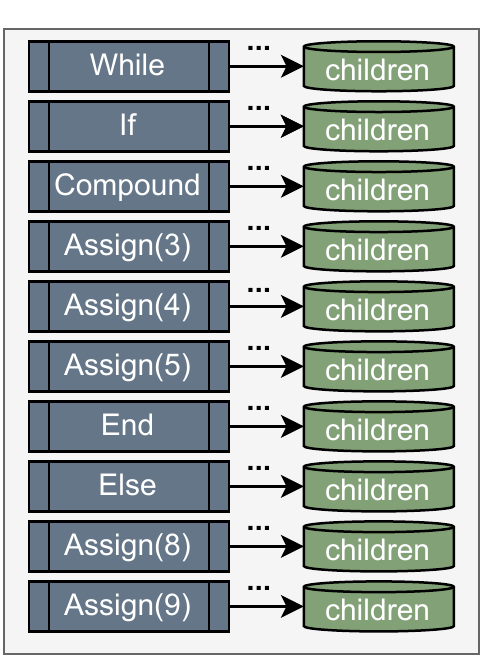}
        {\footnotesize (b) Statement subtree sequence.}
    \end{minipage}
    \caption{Illustration of an example code segment and its corresponding statement subtree sequence. The subtrees are extracted according to the statements in code segment. In subfigure (b), the objects marked with grey represent subtree roots and the objects marked with green represent children of roots. The numbers followed by Assign indicate the line numbers of the code segment.}
    \label{fig:example}
\end{figure}

The main body of Algorithm \ref{alg:split} consists of three procedures. It starts with initializing an empty sequence for storing subtrees (line 10) and then applying a preorder traversal to generate the statement subtree sequence (line 11) that is finally returned (line 12). In function \texttt{preorderTraversal}, it accepts the root node of an AST as input (line 1). When this root node belongs to subtree root identifiers (line 2), we construct a statement subtree based on this root (line 3) and append this statement subtree to the subtree sequence (line 4). Then, for each child node of this root that does not belong to the statement subtree (line 6), we recursively perform \texttt{preorderTraversal} (line 7) to generate the other statement subtrees, i.e., statements inside program branches.

To help understand how the source code is transformed into a statement subtree sequence, we present an example as shown in Figure \ref{fig:example}. We can observe that every subtree in the sequence corresponds to a statement in the code segment, which is obtained by preoreder traversal and is in the order of: While, If, Compound, Assign(3), Assign(4), Assign(5), End, Else, Assign(8), Assign(9). 

\subsection{Gated Recursive Unit over Tree Structures}
Through the statement subtree sequence extracted from the AST, the code naturalness is incorporated. To better capture the syntactical information of the subtrees, we propose gated recursive unit (GRvU) based on previous work \cite{tai2015improved, chen2015gated, chen2015sentence, chen2017improved, kokkinos2017structural}. In contrast, we simplify the complicated components of their mechanisms and make the recursive neural network sufficiently efficient for industrial practice. That is, the position-aware fully connected layer is removed to reduce time complexity and enable full parallelization. 

\begin{figure}
    \centering
    \includegraphics[width=\linewidth, trim=0 5 0 0]{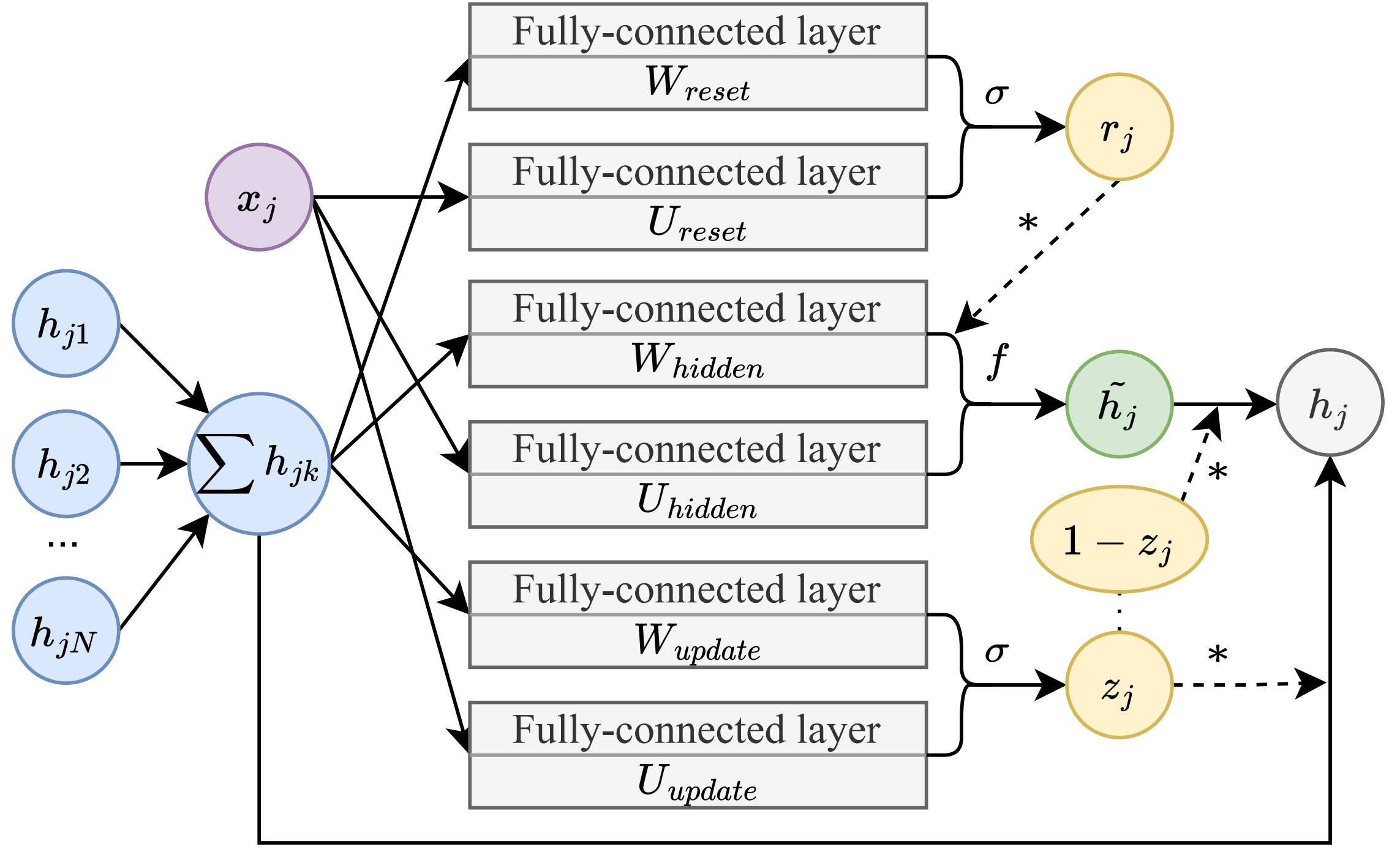}
    \caption{Illustration of the proposed GRvU. GRvU performs bottom-up recursive aggregation to learn the syntactical information. By introducing hidden states and the gating mechanism to the recursive neural network, we are allowed to acquire high-quality subtree representations.}
    \label{fig:GRvU}
\end{figure}

Figure \ref{fig:GRvU} introduces the workflow of the proposed GRvU. GRvU assigns each node of the subtree a hidden state, which records the bottom-up information of it child nodes. We further apply gating mechanism to incorporate the information of hidden states and inputs. Specifically, given a subtree node $j$, its hidden state is the interpolation of the previous calculated hidden states $h_{jk}$ of its $k$-th child out of $N$ total children and the candidate hidden state $\Tilde{h}_j$. It is calculated as follows:
\begin{equation}
    h_j = z_j * \sum_{k = 1}^N h_{jk} + (1 - z_j) * \Tilde{h}_j
\end{equation}
where $z_j$ is the update gate, which keeps a part of the hidden states of children and the other part of its candidate hidden state. The calculation of the update gate is as follows:
\begin{equation}
    z_j = \sigma(U_z * x_j + \sum_{k = 1}^N W_z * h_{jk})
\end{equation}
where $U_z \in \mathbb{R}^{d \times d}$ and $W_z \in \mathbb{R}^{d \times d}$ represent fully-connected layer, and $\sigma$ represent sigmoid function that can map the inputs to the interval from 0 to 1. Besides, we also introduce a reset gate to selectively filter hidden states of children, which is computed as follows:
\begin{equation}
    r_j = \sigma(U_r * x_j + \sum_{k = 1}^N W_r * h_{jk})
\end{equation}
The reset gate is applied to choose important elements from the hidden states of children, which is also activated by the sigmoid function. Combining the input of the current root and gated hidden states of children, we obtain the candidate hidden state of node $j$. It contributes to generating the final hidden state of node $j$. The candidate hidden state $\Tilde{h}_j$ is computed as:
\begin{equation}
    \Tilde{h}_j = f(U_h * x_j + \sum_{k = 1}^N W_h * (h_{jk} * r_j))
\end{equation}
where $f$ represents the hyperbolic tangent function that activates the input. We consider the hidden state of the subtree root as its distributed representation. Therefore, by applying the proposed GRvU over each subtree of the subtree sequence, we can acquire the representation of the subtree sequence:
\begin{equation}
    O = [o_1, o_2, ..., o_M] = \texttt{GRvU}(S)
\end{equation}
where $O$ preserves the effective syntactical information of each statement, which can significantly improve the generalizability of our xASTNN.

\subsection{Gated Recurrent Unit for Subtree Sequence}

We also adopt the well-acknowledged gated recurrent unit (GRtU or GRU) \cite{cho2014properties} to capture the sequential information of subtree sequence, which constitutes the code naturalness together with the aforementioned syntactical information. Given the outputs from GRvU, we adopt a standard bidirectional GRtU to learn the relation between these subtrees. The calculation of GRtU is as follows:

\begin{equation}
    V = [v_1, v_2, ..., v_M] = concat(\texttt{GRtU}(\overrightarrow{O}), \texttt{GRtU}(\overleftarrow{O}))
\end{equation}
where $concat$ represents the concatenation of two vectors, which combines the subtree representations of two directions. Finally, both syntactical information and sequential information have been introduced into the subtree embeddings, thereby ensuring that code naturalness is well modeled. To obtain the final code representation, we feed the enhanced subtree embeddings into a max pooling layer:

\begin{equation}
    c = \max([v_1, v_2, ..., v_M]) = \max(V)
\end{equation}
where $c$ is the code representation produced by our xASTNN. $\max$ is utilized to choose the most important semantics of enhanced subtree representations.

\begin{figure}[t]
    \centering
    \subfigure[Previous approach.]{
    \includegraphics[width=0.45\linewidth]{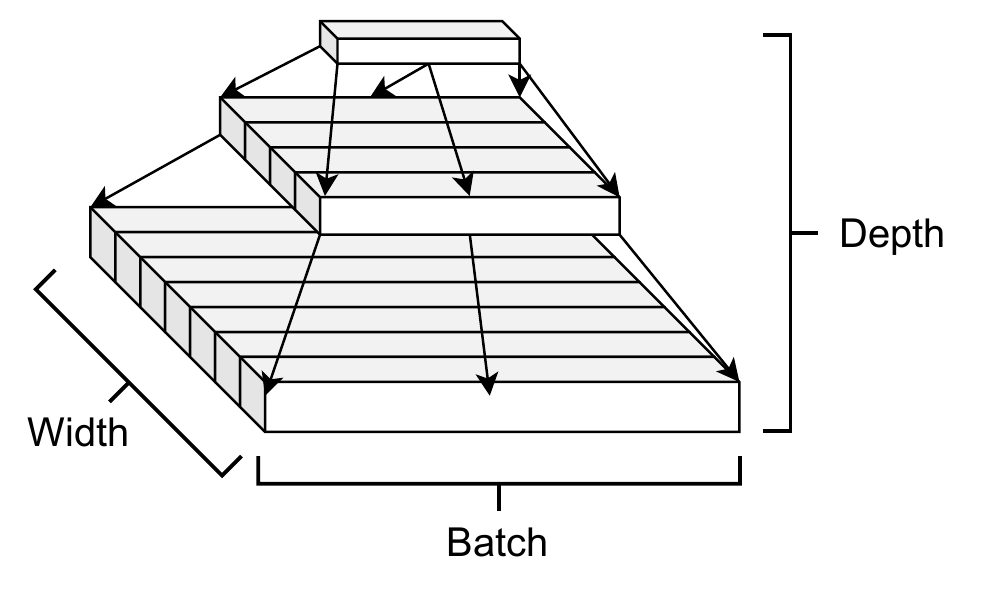}
    }
    \subfigure[Our approach.]{
    \includegraphics[width=0.45\linewidth]{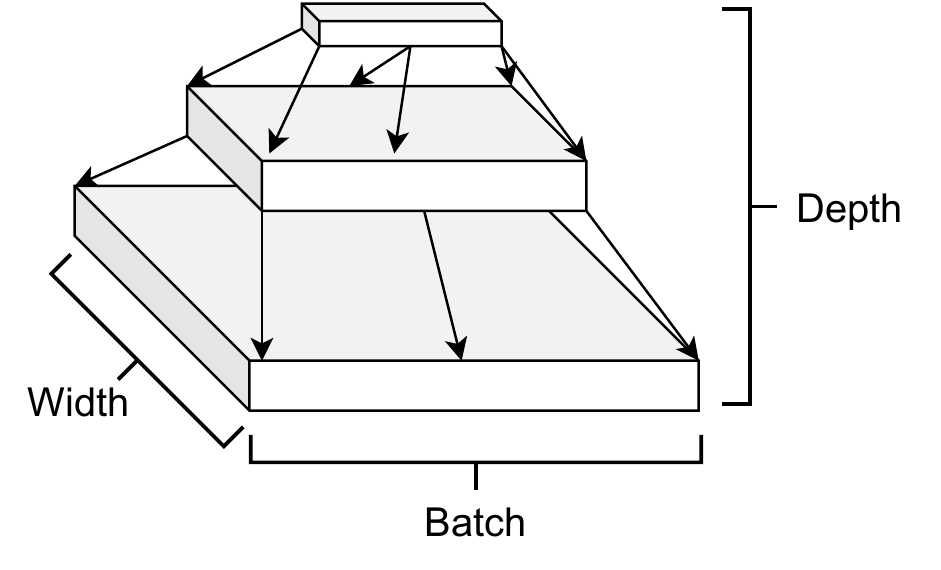}
    }
    \caption{Comparison between previous batching algorithm and our batching algorithm. A cuboid represents one execution of the CPU/GPU. It can be observed that our dynamic batching algorithm supports full parallelism in the width dimension, which accelerates the time complexity of GRvU.}
    \label{fig:batch}
\end{figure}

\subsection{Dynamic Batching Algorithm}
\label{sec:batching}

In order to make our xASTNN competent in industrial practice, we propose an acceleration method named dynamic batching algorithm for GRvU, aiming to greatly improve the efficiency of the proposed approach. This algorithm allows completely parallel computation on subtree nodes of the same depth in a batch of data samples. Previous approach \cite{zhang2019novel} also makes efforts to speed up the recursive network, however, it still suffers from the incomplete parallel computation. Figure \ref{fig:batch} illustrates the comparison between the previous study \cite{zhang2019novel} and ours. Given a batch of data samples, the previous approach \cite{zhang2019novel} processes the children of subtrees one by one. However, our dynamic batching algorithm supports full parallelism even in the width dimension, which is much faster than the previous approach.

\begin{algorithm}[t]
    \SetAlgoLined 
	\caption{Dynamic batching algorithm for GRvU}
        \label{alg:batching}
	\KwIn{List of subtree sequences $\mathcal{S} = [S_1, S_2, ..., S_T]$\;}
	\KwOut{List of subtree embedding sequences $\mathcal{O} = [O_1, O_2, ..., O_T]$\;}
	
	\BlankLine
	\SetKwFunction{FA}{bottomUp}
    \SetKwProg{An}{Function}{}{\KwRet $rootStates$}
    \An{\FA{$subtrees$}}{
        $roots \leftarrow []$ 
        
        $children \leftarrow []$
        
        \For{each \rm{subtree} $s \in subtrees$}{
            append root node $root$ of subtree $s$ to $roots$
            
            \For{each \rm{child subtree} $c \in s$}{
                append child subtree $c$ to $children$
            }
        }
        
        $rootVec \leftarrow \texttt{embed}(roots)$
        
        $childStates \leftarrow \texttt{bottomUp}(children)$
        
        $rootStates \leftarrow \texttt{GRvU}(rootVec, childStates)$
    }
    \BlankLine
    \BlankLine
	
	$lengths \leftarrow []$
	
	$batch \leftarrow []$
	
	\For{each \rm{subtree sequence} $S \in \mathcal{S}$}{
	    append length of $S$ to $lengths$
	    
	    \For{each \rm{subtree} $s \in S$}{
	        append subtree $s$ to $batch$
	    }
	}
	
	$batchVec \leftarrow \texttt{bottomUp}(batch)$

    $\mathcal{O} \leftarrow []$
    
    $shift \leftarrow 0$
    
	\For{each \rm{length} $l \in lengths$}{
	    $O \leftarrow batchVec[shift:shift + l]$
	    
	    $shift \leftarrow shift + l$
	    
	    append subtree embedding sequence $O$ to $\mathcal{O}$
	    
	}
	
	\KwRet $\mathcal{O}$
\end{algorithm}

The workflow of dynamic batching algorithm is shown in Algorithm \ref{alg:batching}. It takes a batch of subtree sequences as input, and outputs the corresponding batch of subtree embedding sequences that contain syntactical information. The main body of Algorithm \ref{alg:batching} starts from line 14. We first initialize two empty lists for storing length of each subtree sequence (line 14) and all subtrees of the batch (line 15), respectively. For each subtree sequence (line 16), we cache the amount of subtrees within the subtree sequence (line 17) and flatten all subtrees into the list $batch$ (lines 18-20). Through performing \texttt{bottomUp} function (lines 1-13), we enhance the subtree embeddings with syntactical information (line 22) and recover them to the original form of a batch (lines 23-29). The recovery procedure starts with the initialization of a list to store final result (line 23) and a record variable (line 24). Then, we extract the corresponding subtree embeddings (lines 26-29) according to length of each subtree sequence (line 25).

The function \texttt{bottomUp} is designed for processing a batch of subtrees simultaneously. It accepts a list of subtrees as input (line 1) and outputs their embeddings learned by GRvU (line 13). This function initializes two lists for storing root (line 2) and children (line 3) of each subtree. For each subtree in the list of subtrees (line 4), we extract the root (line 5) and its children (lines 6-8), which enables the algorithm to process the contents of the same tree depth within a batch of data samples. We apply \texttt{embed} to produce the embeddings of roots (line 10) and \texttt{bottomUp} to acquire the hidden state of subtrees (line 11). At last, the hidden states of current roots can be calculated using \texttt{GRvU} (line 12). This recursion ultimately returns the embeddings of the subtrees that initially provided to the function \texttt{bottomUp} (line 13).

The overall time complexity of dynamic batching algorithm is $O(BLWD)$, where $B$ is the batch size, $L$ is the average length of subtree sequence, $W$ is the average width of subtrees, and $D$ is the average depth of subtrees. It comes from three parts: $O(BL)$ for flattening subtrees, $O(BLWD)$ for bottom-up recursion, and $O(BL)$ for recovering a batch of subtree sequences. Further, given a fact that the AST size $O(A)$ is basically equal to the product of $L$, $W$, and $D$, we can conclude that the overall time complexity of the proposed algorithm is linear to the batch size $B$ and the AST size $A$. Here, the batch size $B$ is a hyperparameter set by developers, but the AST size $L$, $W$, and $D$ is influenced by the data and the granularity to transform AST into subtrees. A smaller $L$ may lead to a larger $W$ and $D$ in the experiments. In our empirical analysis, we found that transforming AST into statement subtrees can advance the generalizability while guaranteeing the efficiency. The rationale might be that statement subtree sequences can well capture the code naturalness and have stable subtree depth. Moreover, the computation time is also affected by a few exceptional circumstances. When some subtrees within a batch have a significantly large depth, our dynamic batching algorithm should conduct extra executions for these subtrees. In practice, we could put the processing of these subtrees together to make the subtree depth as balanced as possible, which we leave for future work.

\subsection{Differences from ASTNN}

Our xASTNN spares no efforts to improve the previous code representation approach for industrial practice. The differences between the previous ASTNN and our xASTNN is as follows.

We adopt gating mechanism throughout the encoding pipeline. Specifically, a child-sum gated recursive unit named GRvU is proposed to encode the syntactical information of statement subtrees. We artificially let the gates position-insensitive, alleviating the computational complexity of space and time for industrial practice. The gates refine the latent information of the code segment, consequently guaranteeing the effectiveness of our xASTNN.

We introduce the gating mechanism also for applicability. The previous approach simply leverages fully-connected layer to capture the information of subtrees, which can be easily trapped in gradient problems. Their feedforward computation is as follows.
\begin{equation}
    h_j = \sigma(W x_j + \sum_{k=1}^N h_{jk} + b)
\end{equation}
When we perform the back propagation for training the model, the gradient of the parameter $W$ in fully-connected layer is as follows.
\begin{equation}
    \frac{\partial J}{\partial W} = \frac{\partial J}{\partial h_0} \frac{\partial h_j}{\partial x_j} + \cdots + \frac{\partial J}{\partial h_j} \frac{\partial h_j}{\partial h_{jk}} \frac{\partial h_{jk}}{\partial h_{jkl}} \cdots \frac{\partial h_{jkl \cdots p}}{\partial h_{jkl \cdots pq}} \frac{\partial h_{jkl \cdots pq}}{ \partial x_{jkl \cdots pq}}
\end{equation}
where $h_{jk}$ denotes the $k$-th child of the node $j$, $h_{jkl}$ denotes the $l$-th child of the node $jk$, and so on. $J$ denotes the loss function. We can observe that the gradient is composed by multiplications of many terms due to the recursive processing of child nodes. The accumulation of these terms may easily lead to a gradient of 0 or infinity, making it hard to train the model. In our scenarios, when the length of code segment is large (e.g., 100), the gradient problems of ASTNN becomes obvious. By contrast, the introduction of gating mechanism in our approach can relieve the gradient problems. The proof can be referred to \cite{hochreiter1997long}. 

Besides, we spare no efforts to optimize the time and space efficiency of our xASTNN. We implement a high-efficient and memory-friendly model compared with the previous approach. That is, our model can still achieve time acceleration and space reduction in the absence of any algorithm. To further accelerate our approach, in this work, we present a dynamic batching algorithm for the processing of subtrees. It adopts more parallel operations and can speed up the model when we feed a batch of data samples.

\section{Evaluation}
\label{sec:evaluation}

We validate the proposed approach in three aspects.

\begin{itemize}
    \item \textbf{How effective is the proposed approach?} In industrial applications, the effectiveness of code representation is of importance. A high-quality representation can lead to better performance in code-related downstream tasks. By comparing metrics of our approach and the baselines with two program comprehension tasks, we evaluate the effectiveness of the proposed approach.
    \item \textbf{What is the efficiency of our approach?} When applying the representation model in practice, the efficiency is a huge challenge. From the perspective of time and space efficiency, we measure the practical usability of each approach.
    \item \textbf{What are the effects of different designs for the proposed xASTNN?} This research aspect plays a key role in the refinement of the model. We explore the performance of alternative designs of our approach by ablating or adjusting some designed modules. The results are organized and analyzed.
\end{itemize}

\begin{table}[t]
    \setlength\tabcolsep{3.25pt}
    \centering
    \caption{Statistics of three datasets. POJ is used for the code classification task; BigCloneBench and OJClone are used for the code clone detection task.}
    \label{tab:dataset}
    \begin{tabular}{lccc}
        \toprule
        \textbf{Dataset} & \textbf{POJ} & \textbf{BigCloneBench} & \textbf{OJClone} \\
        \midrule
        Language & C & Java & C \\
        Task & classification & clone & clone \\
        \midrule
        \# of code segments & 52,000 & 59,688 & 7,500 \\
        \# of categories & 104 & 2 & 2 \\
        \midrule
        Max. tokens & 8,737 & 16,253 & 2,271\\
        Avg. tokens & 245 & 227 & 244\\
        Max. AST depth & 76 & 192 & 60\\
        Avg. AST depth & 13 & 10 & 13\\
        Max. AST nodes & 7,027 & 15,217 & 1,624\\
        Avg. AST nodes & 190 & 206 & 192\\
        \bottomrule
    \end{tabular}
\end{table}

\subsection{Experimental Setup}
\subsubsection{Datasets}

We conduct two downstream tasks including code classification and code clone detection for evaluation. The code classification measures the fundamental ability to comprehend the programs. And the code clone detection measures the ability to compare two code segments. Table \ref{tab:dataset} illustrates the detailed description of used datasets, with the statistics of code segments, categories, tokens, AST depth, and AST nodes.

In the code classification task, we adopt a widely used \cite{mou2016convolutional, zhang2019novel, ben2018neural, han2021comparison, siow2022learning, zhang2020generating, bui2021infercode} public dataset named POJ\footnote{\url{https://sites.google.com/site/treebasedcnn}} \cite{mou2016convolutional} to measure the quality of code representations. This dataset is collected from a pedagogical programming open judge system, which consists of a large number of programming problems. Students submit their source codes as solutions; the judge system will automatically validate the correctness of the solutions. POJ contains 104 programming problems, which are considered as categories predicted by approaches. Each problem contains 500 C programs, which are considered to belong to the same class. We randomly divide the total 52,000 programs into training, validation, and testing sets with a proportion of 3:1:1. 

We exploit two widely used datasets for the code clone detection tasks: BigCloneBench\footnote{\url{https://github.com/clonebench/BigCloneBench}} \cite{svajlenko2014towards} and OJClone, which are also used in \cite{wei2017supervised, zhang2019novel, han2021comparison, zhao2018deepsim, bui2021infercode}. BigCloneBench is a handcrafted dataset that consists of known true and false positive clones. It was built by mining at first and then manually checking clones of ten common functionalities, with 3 judges over 216 hours of manual validation efforts. BigCloneBench is collected from 25,000 systems, covers 10 functionalities including 6,000,000 true clone pairs. Similar to previous work \cite{zhang2019novel, bui2021infercode, wei2017supervised}, we randomly select 100 thousand samples for the convenience of evaluation. We have manually checked that most of the code segments within BigCloneBench are methods, which is different from another code clone detection dataset, OJClone, where code segments are generally functionalities implemented by multiple methods. OJClone derives from POJ automatically. As \cite{wei2017supervised, zhang2019novel, han2021comparison, zhao2018deepsim, bui2021infercode} did, we choose the first 15 programming problems from POJ, which produces $15 \times 500 = 7500$ code segments. In OJClone, the two segments from the same programming problems form a clone pair; otherwise, they belong to a non-clone pair. This will provide us with 28 million clone pairs, making it immensely time-consuming to conduct experiments. Likewise \cite{wei2017supervised, zhang2019novel, bui2021infercode}, we randomly select 50 thousand samples instead. The OJClone dataset is generated completely automatically, without the manual checking of experts. Most of the clone pairs within OJClone are syntactically dissimilar so that the comparative ability of the competing approaches can be well measured.

\subsubsection{Metrics}
In the code classification task, we choose accuracy in test set as the evaluation metric. It represents the proportion that how many data samples are correctly classified. 

In the code clone detection task, we apply precision, recall, and F1 score as the metrics. Precision indicates how many of the predicted clone pairs are really clone pairs. Recall indicates how many of the clone pairs are correctly predicted. F1 score is the harmonic mean of the precision and recall.

\subsubsection{Baselines}

We consider the following approaches as the baselines for code classification. Some of them can also be applied in the code clone detection task.

\begin{itemize}
    \item SVM+N-gram \cite{cortes1995support}: A machine learning-based approach that incorporates SVM and N-gram for code classification.
    \item Transformer \cite{vaswani2017attention}: A recent popular approach that introduces the parallel attention mechanism and deep residual block to encode the tokens. 
    \item CodeBERT \cite{feng2020codebert}: A bimodal pre-training approach for natural language and programming language based on Transformer neural architecture.
    \item TreeLSTM \cite{tai2015improved}: A novel approach that applies LSTM over ASTs to produce code representations.
    \item TBCNN \cite{mou2016convolutional}: A novel approach that applies convolutions over ASTs to produce code representations.
    \item code2vec \cite{alon2019code2vec}: A recent popular code representation approach that extracts AST paths from the AST at first and applies attentional aggregation upon these paths to produce the code representation. 
    \item ASTNN \cite{zhang2019novel}: A novel superior approach that follows a similar pipeline compared with ours, with lower generalizability and serious efficiency problem.
    \item InferCode \cite{bui2021infercode}: A self-supervised pre-training approach, which pre-trains a TBCNN model by predicting subtrees. 
    \item inst2vec \cite{ben2018neural}: A novel approach that first constructs the conteXtual flow graph and applies RNN to produce the code representation.
    \item GGNN \cite{allamanis2017learning}: A well-designed graph neural approach, which first adds edges to ASTs and adopts GGNN \cite{li2015gated} to represent the source code.
    \item GraphCodeBERT \cite{guo2020graphcodebert}: A upgraded version of CodeBERT that considers data flows of programs.
\end{itemize}

We also select three representative approaches designed for code clone detection particularly.

\begin{itemize}
    \item Deckard \cite{jiang2007deckard}: A traditional approach that utilizes subtrees to identify code clones efficiently.
    \item SourcererCC \cite{sajnani2016sourcerercc}: A traditional token-based code clone detection approach, which can detect exact and near-miss clones efficiently.
    \item CDLH \cite{wei2017supervised}: A deep learning-based clone detector that adds a hash function to TreeLSTM for code clone detection.
\end{itemize}

All the experiments are conducted on a 64-bit platform equipped with 12-core Intel(R) i7-12700KF CPU@3.60GHz, 128GB of RAM, and a 24GB RTX 3090 GPU. 

\begin{table}[t]
    \centering
    \caption{Results of code classification.}
    \label{tab:classification}
    \setlength\tabcolsep{8pt}
    \begin{tabular}{llc}
        \toprule
        \textbf{Groups} & \textbf{Approach} & \textbf{Test Accuracy} \\
        \midrule
        \multirow{3}{*}{Token-based} & SVM+N-gram & 0.847 \\
        & Transformer & 0.907\\
        & CodeBERT & 0.975 \\
        \midrule
        \multirow{5}{*}{AST-based} & TreeLSTM & 0.860\\
        & TBCNN & 0.940\\
        & code2vec & 0.913\\
        & ASTNN & 0.981 \\
        & InferCode & 0.980 \\
        \midrule
        \multirow{3}{*}{Graph-based} & inst2vec & 0.948\\
        & GGNN & 0.961\\
        & GraphCodeBERT & 0.982 \\
        \midrule
        Our approach & xASTNN & \textbf{0.985} \\ 
        \bottomrule
    \end{tabular}
\end{table}

\subsection{Effectiveness Assessment with Two Tasks}
\subsubsection{Code Classification}
We exploit the code classification task to measure the fundamental ability of the models to comprehend the programs. In this task, three categories of baselines are considered, including two token-based approaches, five AST-based approaches, and two graph-based approaches. Among them, CodeBERT and InferCode are representative pre-training techniques designed for programming languages. Table \ref{tab:classification} reports the accuracy in test set of competing approaches.

It can be seen that our xASTNN outperforms all baselines, achieving the highest accuracy of 0.985. Its advantages in accuracy over baselines are significant, improving token-based approaches 1.03\% to 16.29\%, AST-based approaches 0.31\% to 14.42\%, and graph-based approaches 0.31\% to 3.6\%. This validates that our approach is effective in learning program semantics and can produce good code representations.

We find that the performance of the token-based approaches differs in test accuracy. SVM+N-gram achieves the lowest accuracy in comparison with other approaches, showing that the use of tokens is not sufficient for code representations. Transformer has a very strong fitting ability among token-based approaches, achieving an accuracy of 0.907. Further, by pre-training the Transformer neural architecture based on a large-scale bimodal task, CodeBERT improves the accuracy of the vanilla Transformer to 0.975.

To improve the comprehension ability of models, many approaches incorporate ASTs for capturing the syntactical information of the source codes. As three representative code feature learning approaches, the accuracy of TreeLSTM, TBCNN, and code2vec is not high, with an accuracy of 0.860, 0.940, and 0.913, respectively. This is because they do not have well-designed modules for processing ASTs and simply incorporate the syntactical information by LSTM, convolutions, and leaf-to-leaf paths. In contrast, the accuracy achieved by ASTNN and InferCode is high. ASTNN is an AST-based approach that introduces code naturalness by statement subtree sequence and InferCode extensively pre-trains their TBCNN model by predicting subtrees.

As for the graph-based approaches, it can be seen that both inst2vec and GGNN achieve relatively high accuracy. They both represent the source codes by constructing flow graphs. The difference is that inst2vec adopts RNN to encode flow graphs while GGNN uses a graph neural network. The pre-trained approach GraphCodeBERT achieves a high accuracy of 0.982, showing that the introduction of syntactical information can improve the performance of the token-based CodeBERT.

\subsubsection{Code Clone Detection}

\begin{table}[t]
    \centering
    \caption{Results of code clone detection. P, R, F1 represents precision, recall, and F1 score, respectively.}
    \label{tab:clone}
    \setlength\tabcolsep{3.25pt}
    \begin{tabular}{lccccccc}
        \toprule
        \multirow{2}{*}{\textbf{Approach}} & \multicolumn{3}{c}{\textbf{BigCloneBench}} & & \multicolumn{3}{c}{\textbf{OJClone}} \\
        \specialrule{0em}{1pt}{1pt}
        \cline{2-4}\cline{6-8}
        \specialrule{0em}{1pt}{1pt}
        & \textbf{P} & \textbf{R} & \textbf{F1} & & \textbf{P} & \textbf{R} & \textbf{F1}\\
        \midrule
        Deckard & 0.93 & 0.02 & 0.03 & & 0.99 & 0.05 & 0.10\\
        SourcererCC & 0.88 & 0.01 & 0.01 & & 0.71 & 0.00 & 0.00\\
        \midrule
        code2vec & 0.83 & 0.91 & 0.87 & & 0.72 & 0.62 & 0.67\\
        TBCNN & 0.89 & 0.95 & 0.92 & & 0.67 & 0.40 & 0.50\\
        CDLH & 0.92 & 0.74 & 0.82 & & 0.47 & 0.73 & 0.57\\
        ASTNN & 0.998 & 0.884 & 0.938 & & 0.989 & 0.927 & 0.955\\
        CodeBERT & 0.947 & 0.934 & 0.941 & & 0.998 &  0.961 & 0.979\\
        GraphCodeBERT & 0.948 & \textbf{0.952} & 0.950 & & \textbf{0.999} & 0.977 & 0.988 \\
        \midrule
        xASTNN & \textbf{0.999} & 0.935 & \textbf{0.966} & & 0.997 & \textbf{0.987} & \textbf{0.992} \\
        \bottomrule
    \end{tabular}
\end{table}

To sufficiently evaluate the effectiveness of approaches, we also introduce code clone detection, a task that is widely used in code refactoring and vulnerability detection. We consider two lines of approaches as baselines, including code representation approaches and specialized clone detectors. The results are shown in Table \ref{tab:clone}.

Comparing with eight baselines, we can find that the effectiveness of our xASTNN is the most superior on two datasets. xASTNN achieves an F1 score of 0.966 on BigCloneBench and 0.992 on OJClone, improving the leading baseline GraphCodeBERT by 1.68\% on BigCloneBench and 0.40\% on OJClone. 

Two traditional approaches, Deckard and SourcererCC, are unable to handle syntactically dissimilar code clones, as revealed in their recall. But their precision is high, which illustrates that they can precisely distinguish syntactically similar code clones. As for deep learning-based approaches, we note that code2vec, TBCNN, and CDLH all achieve relatively high F1 score compared with the traditional approaches. The advantages of these approaches come from the deep learning features. Nevertheless, it can be observed that ASTNN, CodeBERT, and GraphCodeBERT have a remarkable performance in this task. This shows that their fitting ability is superior compared to the other deep learning-based approaches. An interesting phenomenon is that ASTNN, CodeBERT, and GraphCodeBERT have high precision and low recall in most cases, suggesting that they tend to make confident predictions. 

\begin{center}
\begin{tcolorbox}[colback=gray!10,
                  colframe=black,
                  width=8.5cm,
                  arc=1mm, auto outer arc,
                  boxrule=0.75pt,
                 ]
\textbf{Conclusion 1:} Through evaluating our approach and the baselines in two code comprehension tasks, we can find that our xASTNN is very generalized and effective.
\end{tcolorbox}
\end{center}

\begin{figure}[t]
    \centering
    \includegraphics[width=\linewidth]{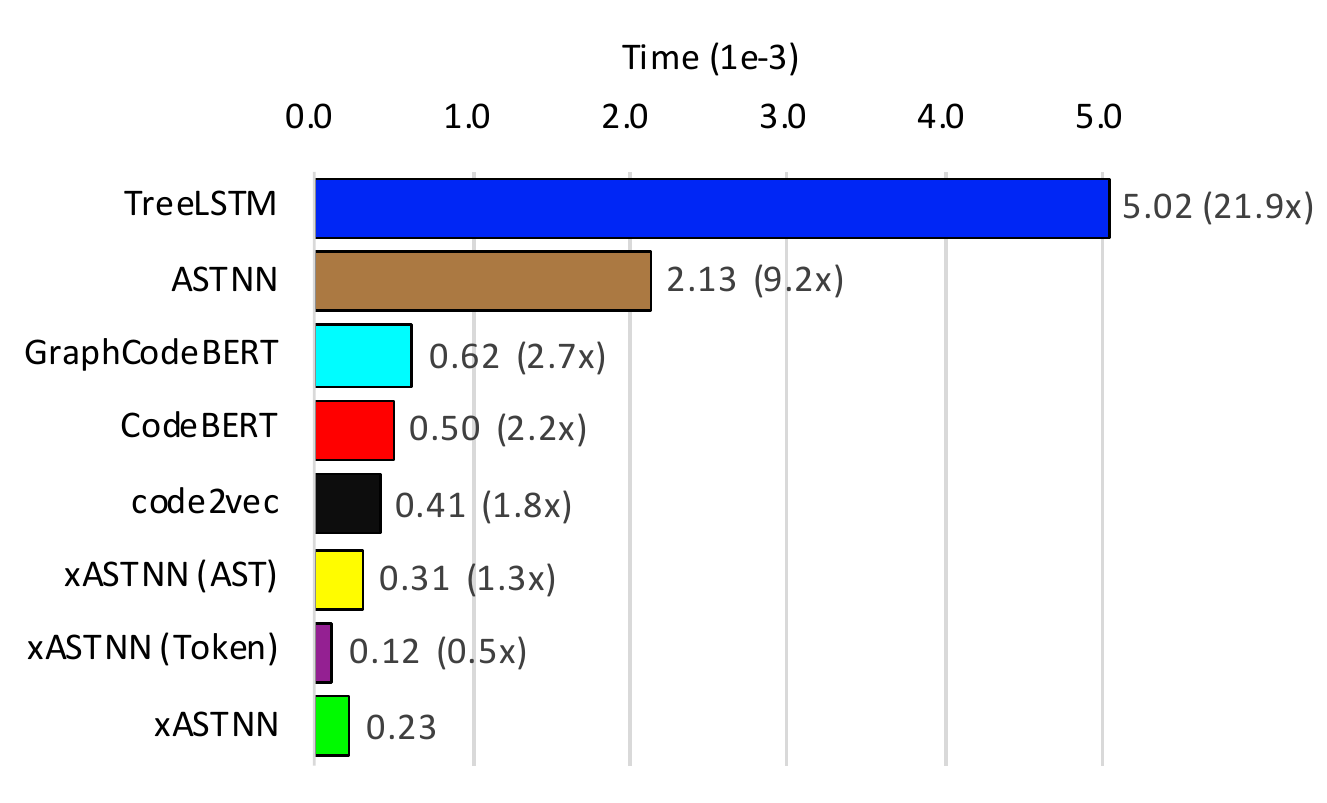}
    \caption{Average computation time of models for representing one code segment. Each approach is tagged with the time (left) and the speedup ratio achieved by our xASTNN (right).}
    \label{fig:efficiency1}
\end{figure}

\subsection{Efficiency of Models}
To evaluate the time efficiency of models, we measure their computation time from the acceptance of a code segment to the output of its representation. The reciprocal of this metric is known as prediction rate, which reflects the number of data samples a model can process in one second. This experiment is conducted on a broad corpus of real-world programming languages. A batch of data samples is fed into the code representation model and the average computation time for representing one code segment is reported.

Figure \ref{fig:efficiency1} shows the computation time of baselines along with our approach and its variants (batch size is set to 64). It can be observed that our approach costs 0.23 ms to represent a code segment, faster than all the baselines. It accelerates code2vec 1.8$\times$, CodeBERT 2.2$\times$, GraphCodeBERT 2.7$\times$, ASTNN 9.3$\times$, and TreeLSTM 21.9$\times$. 

The computation time of baselines is high, ranging from 0.41 ms to 5.02 ms. code2vec is fast because it represents source code by parallelly encoding a bag of AST paths. CodeBERT parallelly processes tokens of source code and GraphCodeBERT adds the processing of data flows to CodeBERT. The computation time of ASTNN and TreeLSTM is long. The reason is that they involve syntactical information processing but no efficient algorithms are particularly designed. 

Two variants with different subtree granularity perform differently in computation time. Here, we consider the two most special cases, i.e., the largest subtree based on the whole AST named xASTNN (AST) and the smallest subtree based on all AST nodes named xASTNN (token). It can be noted that when the size of subtree gets smaller, the computation time for representing one code segment gets smaller. The reason is that smaller subtrees need fewer recursive operations and these operations are exactly taking up a large portion of the overall computation time.

\begin{figure}
    \centering
    \subfigure[Time v.s. Batch size]{
    \label{fig:efficiency21}
    \includegraphics[width=0.475\linewidth]{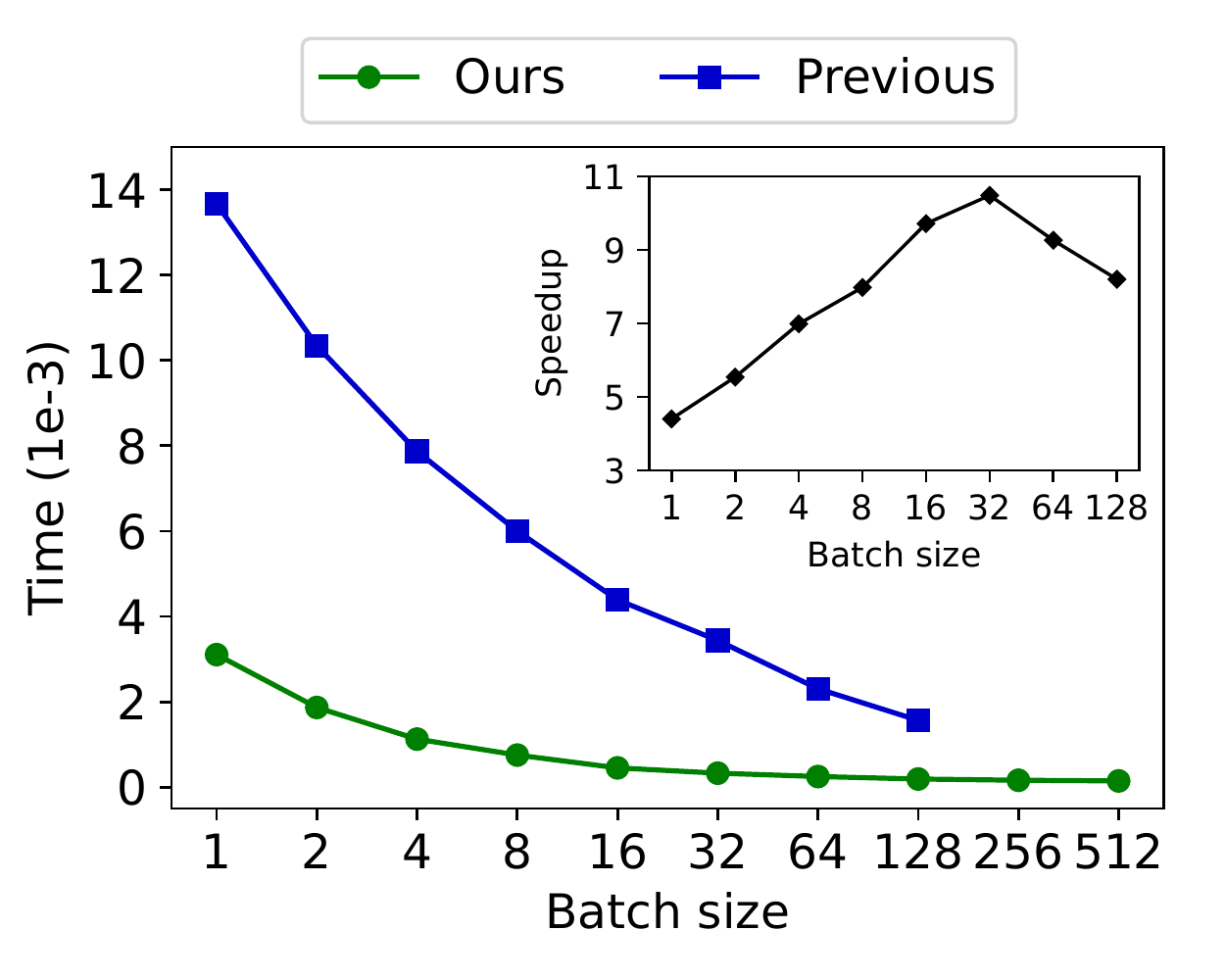}
    }
    \subfigure[GPU usage v.s. Batch size]{
    \label{fig:efficiency22}
    \includegraphics[width=0.475\linewidth]{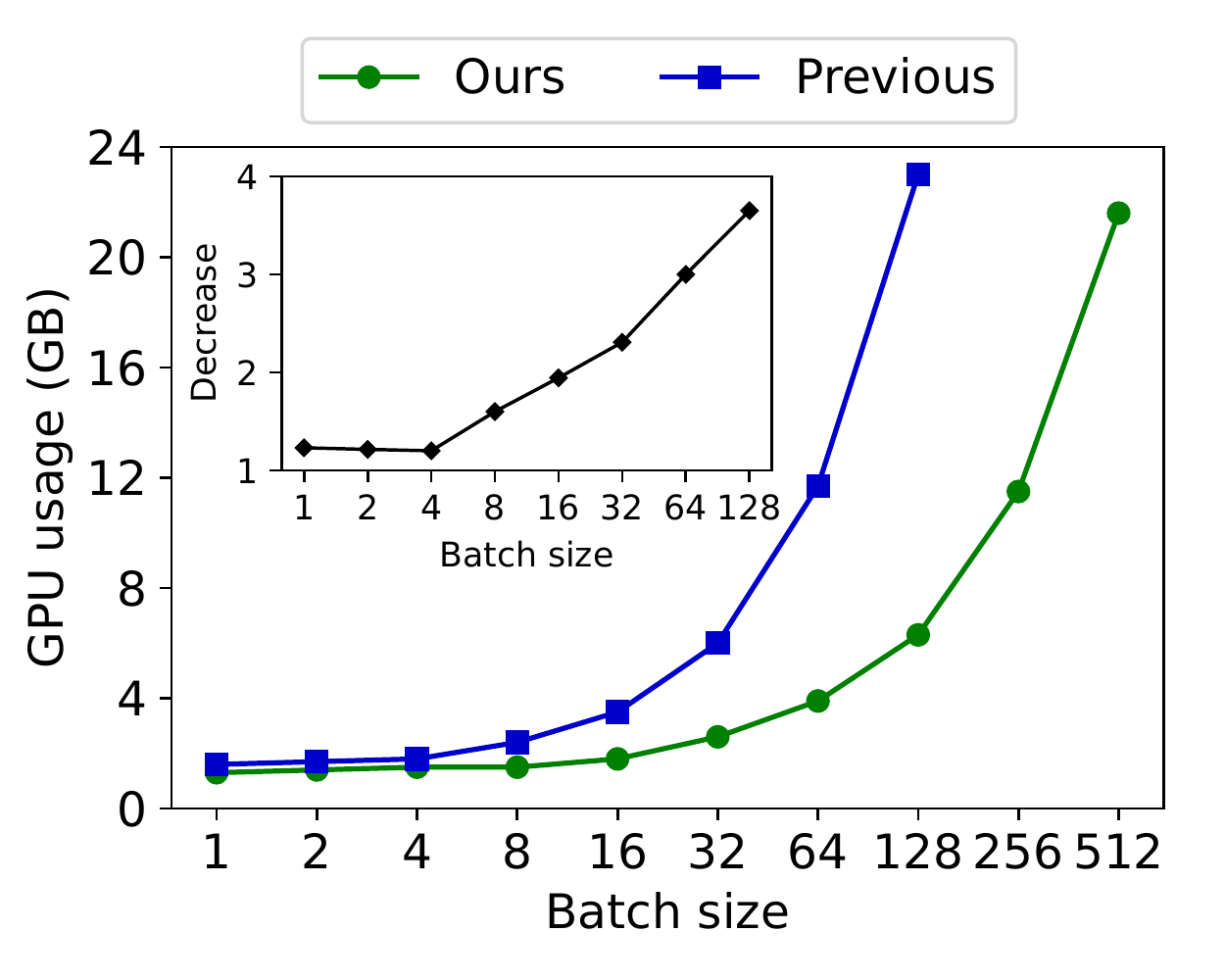}
    }
    \caption{Correlation of time and space efficiency with batch size. The subfigures indicate how many times our approach has improved against the previous approach.}
    \label{fig:efficiency2}
\end{figure}

In addition to the experiment of average computation time, we conduct another experiment to measure the effect of batch size on the time and space efficiency of the previous approach and our approach, aiming to validate the performance of the proposed dynamic batching algorithm. Figure \ref{fig:efficiency2} illustrates the results. Note that when the batch size is larger than 128, the previous approach ASTNN will suffer from a memory overflow problem. Hence, we do not report the corresponding results.

In the aspect of time efficiency, we can observe that our approach is faster than the previous approach at all batch sizes, indicating the superiority of our dynamic batching algorithm. As the batch size increases, the computation times of two approaches both decrease. However, their rates of decline are different. We note that the speedup ratio first increases and then decreases as the batch size goes. The peak value of speedup ratio (i.e., 10.5$\times$) occurs when the batch size is around 32. The rationale for this phenomenon can be inferred from their time complexity (see Figure \ref{fig:batch}). With the increase of batch size, many subtrees of different depths will appear in the batch and dominates the time complexity. This requires our approach to cost plenty of time for these special subtrees, thus resulting in the attenuation of the speedup ratio. Theoretically, if similar-sized subtrees are processed together, this problem will be significantly alleviated. We leave it for future work.

In the aspect of space efficiency, we can see that our approach uses less memory at all batch sizes compared with the previous approach. The GPU usage of both two approaches increases quickly as the batch size increases. When the batch size is equal to 1, our approach still has the advantage in GPU usage, showing its superiority in the processing of single data sample. Similar phenomenon can be found in terms of time efficiency. This is because we optimize the implementation of our xASTNN to make it as lightweight as possible for industrial practice. Hence, when the batch size is set to be large, the improvement in GPU usage of our approach becomes more pronounced. Additionally, an interesting phenomenon is that the GPU usage does not change obviously when the batch size is less than 4. The model size is the main influencing factor for GPU usage at this point.

\begin{center}
\begin{tcolorbox}[colback=gray!10,
                  colframe=black,
                  width=8.5cm,
                  arc=1mm, auto outer arc,
                  boxrule=0.75pt,
                 ]
\textbf{Conclusion 2:} The efficiency of our xASTNN is promising, which is reflected in both time and space. The computation time on the order of 10$^{-4}$ seconds allows our approach to be used in a wide range of industrial applications.
\end{tcolorbox}
\end{center}

\begin{table}[t]
    \centering
    \setlength\tabcolsep{4.5pt}
    \caption{Results of alternative designs of our xASTNN.}
    \begin{tabular}{llc}
    \toprule
    \textbf{Alternatives} & \textbf{Approach} & \textbf{Test Accuracy}\\
    \midrule
    Our approach & xASTNN & 0.985\\
    \midrule
    Subtree & Program Subtree & 0.952\\
    granularity & Token Subtree & 0.971\\
    \midrule
    Syntactical & Removing GRvU & 0.827\\
    information & RvNN Instead of GRvU & 0.981\\
    \midrule
    Sequential & Removing GRtU & 0.976\\
    information & RtNN Instead of GRtU & 0.981\\
    \bottomrule        
    \end{tabular}
    \label{tab:alternative}
\end{table}

\subsection{Effect of Alternative Designs}
\label{sec:alternative}

In this experiment, we investigate the effect of alternative designs for the proposed approach xASTNN, aiming to give explanations about the efficacy of each designed components. We consider POJ as the representative dataset, on which all the results are produced.

At first, we evaluate the alternative designs from the following three perspectives: program subtree or token subtree from the perspective of subtree granularity, removing or replacing GRvU from the perspective of syntactical information, and removing or replacing GRtU from the perspective of sequential information. The performance of these variants are reported in Table \ref{tab:alternative}.

It can be observed that our approach outperforms all the variants, showing that the current design of our approach is effective. When we adjust the subtree granularity, the accuracy decreases by 0.033 to 0.952 for program subtree and by 0.014 to 0.971 for token subtree. This result exhibits that an appropriate subtree granularity is of importance in capturing code naturalness. If the granularity of the subtree is too large or too small, the syntactical or the sequential information will be lost during the modeling process.

When we modify the encoder for syntactical or sequential information, the performance of these variants are different. When we replacing GRvU with RvNN or replacing GRtU with RtNN, the accuracy drops by 0.004 to 0.981. Nevertheless, if we remove one of the encoders, the accuracy will decline drastically, resulting in a decrease in accuracy of 0.158 or 0.009 respectively. This demonstrates that both syntactical information and sequential information introduced by our xASTNN plays a key role in code representations.

\begin{table}[t]
    \centering
    \caption{Correlation of GPU usage, Time, and Accuracy with embedding dimension for our xASTNN. }
    \label{tab:dimension}
    \setlength\tabcolsep{7pt}
    \begin{tabular}{cccc}
        \toprule
        \textbf{Dimension} & \textbf{GPU (GB)} & \textbf{Time (ms)} & \textbf{Accuracy}\\
        \midrule
        2 & 1.4 & 0.23 & 0.065\\
        4 & 1.4 & 0.23 & 0.229\\
        8 & 1.5 & 0.23 & 0.704\\
        16 & 1.7 & 0.23 & 0.904\\
        32 & 1.9 & 0.24 & 0.958\\
        64 & 2.9 & 0.24 & 0.976\\
        128 & 4.5 & 0.24 & 0.984\\
        256 & 5.3 & 0.25 & 0.984\\
        512 & 9.3 & 0.28 & 0.985\\
        1024 & 17.2 & 0.37 & 0.985\\
        \bottomrule
    \end{tabular}
\end{table}

In addition, we introduce another alternative design, namely xASTNN of different sizes, to investigate the effect of model size on performance. We vary the embedding dimension in xASTNN and the results are reported in Table \ref{tab:dimension}. It shows the correlation of GPU usage, time, and accuracy with embedding dimension. 

From the results, it can be noted that the GPU usage quickly increases as the dimension increases, from 1.4 GB to 9.3 GB. The GPU usage grows slowly at first, which is because the running buffer takes up a large amount of space. When the embedding dimension reaches to a certain degree (e.g., 64), its impact on GPU usage becomes significant. As for the computation time, a similar phenomenon appears when the embedding dimension reaches 512. The accuracy is strongly influenced by the embedding dimension. It starts at 0.065, increases rapidly to over 0.976, and converges. This demonstrates that the quality of code representations is highly related to the model size. Therefore, we should consider balancing space, time, and accuracy in practice.

\begin{center}
\begin{tcolorbox}[colback=gray!10,
                  colframe=black,
                  width=8.5cm,
                  arc=1mm, auto outer arc,
                  boxrule=0.75pt,
                 ]
\textbf{Conclusion 3:} The design of our xASTNN is reasonable. Each module of our approach has a different effect on its performance, requiring the developers to carefully tune the parameters according to the production requirements.
\end{tcolorbox}
\end{center}

\section{Threats to Validity}
\label{sec:threats}

In this section, we discuss the threats to our work. The first limitation is that some of our experiments are not based on real-world programming language corpus. For convenience of extensive evaluations on the performance of baselines and our approach, we choose to adopt the widely-used benchmark datasets instead of our non-public data. Additionally, recent study \cite{krinke2022bigclonebench} suggests that BigCloneBench are considered harmful for evaluating machine learning approaches. Despite these defects, we believe that our experiments are still worthy to be used as a reference for the superiority of our xASTNN. The second limitation is that the insufficient investigation of the robustness of the competing approaches. The results reported in this work are in the common form of the average or maximum performance. Nevertheless, the stability of model performance is not measured. This measurement can illustrate whether a model has performance jitter and makes abnormal predictions. We leave this to future work.

\section{Lessons Learned}
\label{sec:lessons_learned}

From the work of developing code representation for industrial practice, we have learned three significant lessons: 

\textbf{Making the code representation model applicable to various scenarios is an important issue.} Developing high-quality source code representations have aroused many interests recently. Some approaches borrow complicated characteristics of programming languages to improve the effectiveness of their models, making them hard to be applied in practice. Besides, an inappropriate model design and implementation can also introduces many practical problems, such as memory overflow, runtime delay, and difficult objective fitting.

\textbf{Reaching a trade-off between effectiveness and efficiency is very high priority for industrial practice.} In different industrial scenarios, the application of code representation model may encounter different computing environments and business requirements. Our experiment results show that the effectiveness of our approach can be improved by increasing the model size, while sacrificing the time and space efficiency. Hence, adjusting the model in terms of the actual needs is advisable in industry.

\textbf{Unusual data inputs have a dramatic impact on the model performance.} There are few extremely long or short code segments in practice, which are often deleted by most of academic experiments. These data samples can easily lead to gradient vanishing or explosion problems. Moreover, a batch of size-unbalanced data samples costs more time to process. The approach should spare computing resources for those special data samples. Hence, if we do not pay attention to these unusual data in industry, the performance of the model will be weakened.

\section{Conclusion}
\label{sec:conclusion}

In this paper, we propose an eXtreme AST-based neural network named xASTNN for producing code representations in industry. The design of xASTNN concentrates on unlocking the potential of AST-based neural network. Our approach is completely based on common ASTs, which is easily accessed by AST parsers. Besides, we introduce techniques such as gating mechanism and dynamic batching algorithm to advance the performance, reduce the computation time, and alleviate the gradient problems. Extensive experiments on two code comprehension tasks have been conducted to demonstrate the effectiveness and efficiency of our xASTNN. According to the results, we can see that our xASTNN outperforms the state-of-the-art while achieving an acceleration of over 10$\times$ than the previous approach ASTNN. Therefore, the proposed approach is lightweight, effective, and efficient, with the promising possibility of being applied in a wide range of industrial applications.

\begin{acks}
This research was supported in part by the National Key Research and Development Program of China (No. 2020YFB1707700), the National Natural Science Foundation of China (No. 92267203, No. 62076146, No. 62021002, No. U20A6003, No. U19A2062, No. U1911401, and No.6212780016), and the Industrial Technology Infrastructure Public Service Platform Project ``Public Service Platform for Urban Rail Transit Equipment Signal System Testing and Safety Evaluation'' (No. 2022-233-225), Ministry of Industry and Information Technology of China.
\end{acks}

\bibliographystyle{ACM-Reference-Format}
\bibliography{ref}

\end{document}